\documentclass[useAMS,usenatbib]{mn2e}

\usepackage[dvips]{graphicx}
\input{epsf}

\title[Dark matter chaos in the Solar System]
{Dark matter chaos in the Solar System}
\author[J. Lages and D. L. Shepelyansky]{
J. Lages$^{1}$\thanks{E-mail:
jose.lages(at)utinam.cnrs.fr (JS);  dima(at)irsamc.ups-tlse.fr (DLS)}
 and D. L. Shepelyansky$^{2}$
\\
$^{1}$ Institut UTINAM, Observatoire des Sciences de l'Univers THETA,
CNRS \& Universit\'e de Franche-Comt\'e, 25030 Besan\c{c}on, France\\
$^{2}$Laboratoire de Physique Th\'eorique du CNRS, IRSAMC, 
Universit\'e de Toulouse, UPS, F-31062 Toulouse, France}
\begin{document}

\date{Accepted 2012 December XX. Received 2012 November 6; in original form 2012 November 6}

\pagerange{\pageref{firstpage}--\pageref{lastpage}} \pubyear{2012}

\maketitle

\label{firstpage}

\begin{abstract}
We study the capture of galactic dark matter particles in the Solar System
produced by rotation of Jupiter. It is shown that the capture
cross-section is much larger than the area of Jupiter orbit
being inversely diverging at small particle energy. We show that the dynamics
of captured particles is chaotic and is well described by a simple
symplectic dark map. This dark map description allows to simulate the 
scattering and dynamics of $10^{14}$ dark matter 
particles during the life time of the Solar System
and to determine dark matter density profile 
as a function of distance from the Sun.
The mass of captured dark matter in the radius of Neptune orbit is
estimated to be $2 \cdot 10^{15} g$.
The radial density of captured dark matter is found to be 
approximately constant behind Jupiter orbit being similar to the density profile
found in galaxies.
\end{abstract}

\begin{keywords}
dark matter, Solar System, Hamiltonian chaos, dynamical maps
\end{keywords}

\section{Introduction}

A galactic wind of dark matter particles (DMP) (see e.g. \cite{bertone2005})
flies through the Solar System and a part of it becomes
captured  due rotation of planets around the Sun.
The capture process, dominated by Jupiter, is related to the nontrivial
aspects of the restricted three-body problem 
(see e.g. \cite{3body}).
We demonstrate that this process is described by a simple 
dynamical symplectic map  (see e.g. \cite{chirikov1979,lichtenberg}) which
allows to perform extensive numerical simulations
of DMP capture. Our studies show  that the capture cross-section 
is much larger than the area of Jupiter orbit
being diverging as an inverse square of DPM velocity
in agreement with recent analytical estimates by
\cite{khriplovich2009}. 

The dynamical map analysis
allows to simulate DMP capture and ejection on
the whole life time scale of the Solar System
for $10^{14}$ DMP being more efficient than
the direct simulations of DPM dynamics by \cite{peter2009}.
Our approach provides a DMP
density distribution in the Solar System
with other features of dynamics at present time
after 4.5 billion years evolution of the Solar System.
This  DMP distribution 
is similar to those found in  galaxies 
by \cite{rubin}.
The dynamics of DMP is shown to be chaotic having certain 
similarities with a chaotic  comet motion in the Solar System
discussed by \cite{petrosky,halley,tremaine1989,dvorak,tremaine1999}. 

Following \cite{bertone2005} we assume 
that in a vicinity of the Solar System (SS)
the velocity distribution of DMP has a Maxwell form 
$f(v) dv = \sqrt{54/\pi} v^2/u^3 \exp(-3v^2/2u^2) dv$
with the average module velocity $u \approx 220 km/s$.
During a scattering of DMP on the Sun its rescaled total energy
$w=-2E/m_d v_p^2$ is changed due to planetary rotation.
The main contribution is given by Jupiter,
as discussed by \cite{halley},
and hence we base our studies on the case of one planet
measuring DMP parameters in units of
planet radius $r_p$ and velocity $v_p$ taken as unity,
DMP mass $m_d=1$.
The studies of comet dynamics by \cite{petrosky,halley,tremaine1989}
in SS with one rotating planet
show that it is well described by
a symplectic map and thus a DMP dynamics
over an extended orbit is also described by that type of map.

\section{Dark map description}
This dark map  has a form similar to the Halley map 
(see \cite{halley}):
\begin{equation}
\label{eq1}
w_{n+1}=w_n+F(x_n) \; , \;\; x_{n+1}=x_n+w^{-3/2}_{n+1} \; ,
\end{equation}
where $x_n= t_n/T_p \; (mod 1)$ is given by time $t_n$ 
taken at the moment of DMP $n-th$ passage through perihelion,
$T_p$ is the planet period.
The second equation in (\ref{eq1})
follows from the Kepler law for the DMP orbital period.
The amplitude of kick function $F(x)$ is proportional 
to the ratio of planet mass $m_p$ to the Sun mass $M_S$ 
($F \sim m_p/M_s$) (see \cite{petrosky,halley}). 
Its shape depends on DMP perihelion distance $q$,
inclination angle $\theta$ between the planetary plane $(x,y)$
and DMP plane, and perihelion orientation angle $\varphi$.
However, $F(x)$ is independent of
energy $w$  for $1/|w| \gg r_p=1$. 
Thus the dark map describes DMP
dynamics for bounded and unbounded energies as well as its
capture process corresponding to a transition from 
positive $w <0$ to negative energies  $w>0$.

\begin{figure}
\begin{center}
\includegraphics[width=0.48\textwidth]{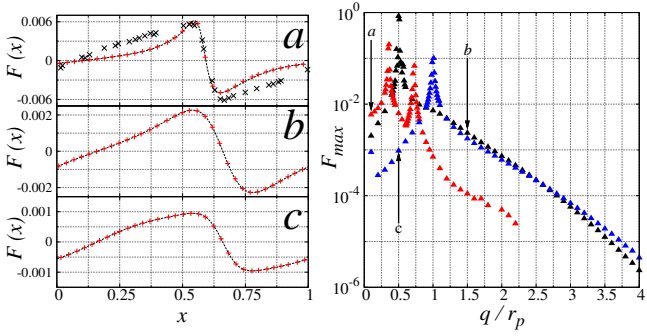}
\end{center}
\caption{ {\em Left panel}: dependence of kick
function $F(x)$ on Jupiter phase
$x$ for DMP orbit parameters shown by pluses:
$a)$ $q=0.11$, $\theta=2.83$, $\varphi=1.95$
of the Halley comet case; here crosses show data
for the Halley comet with all SS planets taken from Fig.1 of 
{\protect \cite{halley}};
$b)$ $q=1.5$, $\theta=0.7$, $\varphi=\pi/2$;
$c)$ $q=0.5$, $\theta=\pi/2$, $\varphi=0$; curves show fit functions of
numerical data marked by pluses.
{\em Right panel}:  dependence of maximal amplitude
$F_{max}$ on $q$ for  $a,b,c$ cases of left panel.
\label{fig1}}
\end{figure}

Our direct numerical
simulations of Newton equations
confirms this map description by $F$-function
as it is shown in Fig.~\ref{fig1} for
various values of $q, \theta, \varphi$,
including the Halley comet case analyzed by \cite{halley}. 
In agreement 
with the theory of \cite{petrosky}
the maximum $F_{max}$ drops exponentially 
for $q \gg r_p$ so that only DMP with
$q < 2 r_p$ can be effectively captured.
At $q \gg r_p$ we find $F \sim \sin 2\pi x$
in agreement with results of \cite{petrosky}.
The visible peaks in $F_{max}$
correspond to close encounters between DPM and planet
happening at rather specific angles for $q \leq r_p$.
We will see later that such events give a small
contribution in the capture cross-section $\sigma$.
In fact, $F-$function contribution comes from 
encounter distances of the order of $r_p$
thus being much larger than the radius of the planet body $r_b$.
This analytical result of \cite{petrosky,halley,khriplovich2009,rydberg}
is in agreement with the detailed numerical simulations by \cite{peter2009}
invalidating previous numerical studies of
\cite{gould2001,edsjo2004}
which considered contributions only from $r_b$ scale.

Finally, we note that the dark map gives an efficient but 
approximate description. For the exact dynamics there is a slow
variation of DMP orbital momentum $\ell$ and $q=\ell^2/(2r_pv_p^2)$
and angles $\theta, \varphi$ (see \cite{dvorak}). 
However, the rate of these variations is rather slow
being proportional to $m_p/M_S$ and does not affect significantly
the chaotic diffusion in energy. Also numerical simulations of 
DMP dynamics by \cite{peter2009}  point on a small global variation of $q$.
A similar situation
appears in a microwave ionization of Rydberg atoms
where it is known that  the Kepler map in energy
gives a good description of ionization process of 3D-atoms
as discussed by  \cite{rydberg}. Also the DPM flow $f(v)dv$ performs
an averaging over all $\ell, \theta, \varphi$ values
and hence  a variation of these parameters is averaged out.

\section{Capture cross-section}
\begin{figure}
\begin{center}
\includegraphics[width=0.48\textwidth]{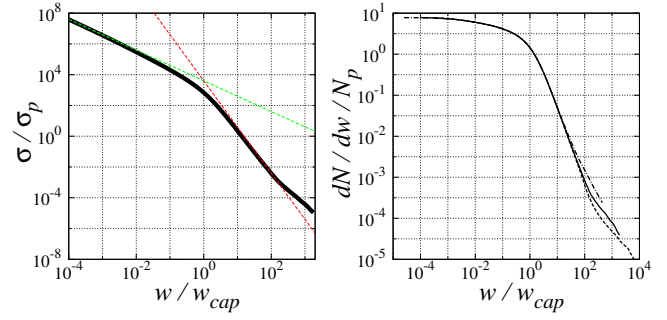}
\end{center}
\vglue -0.4cm
\caption{{\em Left panel}: Dependence of capture
cross-section $\sigma$ for Jupiter on DMP energy $w$;
dashed lines show dependence $\sigma \propto 1/|w|; 1/w^2$.
{\em Right panel}: Dependence of rescaled 
captured number of DMP on energy $w$ for Jupiter,
Saturn and a model planet with $m_p/M_S=0.004$
(full, dashed, dot-dashed curves respectively).
\label{fig2}}
\end{figure}

In a scattering problem at infinity we have $\ell^2 = r_d ^2 v_p^2 |w|$
with the impact scattering distance $r_d^2 = 2 q r_p /|w|$. Hence,
the capture cross-section at energy $|w|$ is 
$\sigma(w)/\sigma_p = (\pi^2 r_p |w|)^{-1} 
\int_0^{2\pi} d \theta \int_0^\pi d\varphi \int_0^{\infty} dq h(q,\theta,\varphi)$,
where $h$ is a fraction of DMP captured
after one map iteration from $w<0$ to $w>0$, 
given by an interval length  inside $F(x)$ envelope 
at $|w|=const$, $\sigma_p=\pi r_p^2$.
This fraction is determined from numerically computed
$F(x)$, as those shown in Fig.~\ref{fig1},
via a continuous fit spline of function $F(x)$.
Using a grid with up to $N_g=10^5$ points in $(q,\theta,\varphi)$
volume we perform a Monte Carlo integration 
which gives $\sigma(w)$ as a function of energy $w$
for the case of Jupiter where the main contribution
is given by $|w| \sim w_{cap}  = m_pv_p^2/M_S \approx 10^{-3}$.

The dependence $\sigma(w)/\sigma_p$ is shown in Fig.~\ref{fig2}.
For $|w|<w_{cap}$ we find 
$\sigma/\sigma_p \approx \pi M_S w_{cap}/m_p |w|$ in  agreement
with analytical estimates by \cite{khriplovich2009},
for $|w| > w_{cap}$ we have 
$\sigma/\sigma_p \approx \pi M_S w_{cap}^2/(m_p w^2)$.
The later regime describes contribution of close
encounters which has a rapid decrease of $\sigma$
and hence gives a small contribution in the capture process.
This conclusion is confirmed by the analysis of
the differential number of captured DMP per time unit  
$d N =  \sigma(w) n_g v_p^2 f(w) d |w| /2$. Here $n_g$
is a galactic DMP density with a corresponding mass
density $\rho_g =m_d n_g \sim 4 \cdot 10^{-25} g/cm^3$ \
(see \cite{bertone2005})
and $f(w)$ is the velocity distribution function given above
with $|w|=v^2/v_p^2$ at infinity.
A number of DMP crossing the planet
orbit area per time unit is
$N_p= \int_0^{1} n_g \sigma_p v_p^2 f(w) d|w|/2$. 

The dependence of
$dN/N_p dw$ on $|w|=v^2/v_p^2$, presented in Fig.~\ref{fig2},
drops quadratically  for $|w| > w_{cap}$ showing that the contribution
of close encounters is small. We note that $dN/N_p dw$
depends only on the ratio $w/w_{cap}$ that is confirmed 
by additional data obtained for Saturn and
a model planet in Fig.~\ref{fig2}.  
 As a result the total number of captured particles is
$N \propto m_p M_S$ in agreement with results of \cite{khriplovich2009}.

\section{Chaotic dynamics}
\begin{figure}
\vskip 0.9cm
\begin{center}
\includegraphics[width=0.48\textwidth]{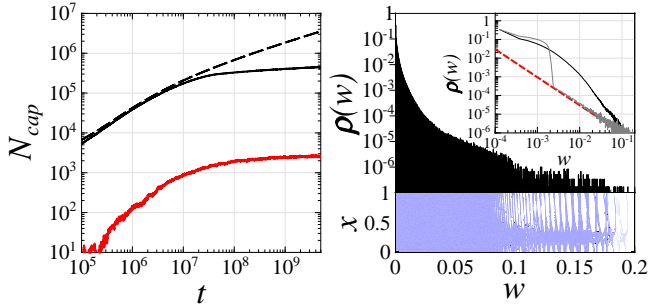}
\end{center}
\vglue -0.4cm
\caption{{\em Left panel}: The number $N_{cap}$  
of captured DMP, as a function of time $t$ in years, 
for  energy range 
$w >0$ (dashed curve), $w >4 \cdot 10^{-5}$ corresponding to
half distance between Sun and 
Alpha Centauri System (black curve), $w >1/20$
corresponding to $r<100 AU$
(red curve); DMP are injected at constant flow $f(v)$ at all
angles.
{\em Right panel}: Top part shows density distribution 
$\rho(w) \propto dN/dw$ in energy
at time $t_S$
for DMP injection at all parameters $q,\theta,\varphi$
(normalized as $\int_0^\infty \rho dw=1$),
bottom part shows the Poincar\'e section of the dark map
for DMP injection at fixed parameters $q,\theta,\varphi$
of Fig. 1b; inset shows 
density distribution of captured DMP in $w$ in log-log scale
for  parameters of Fig. 1b (gray curve) and
orbits of the main right panel injected at all parameters
(black curve), a dashed line shows a slope -3/2.
\label{fig3}}
\end{figure}

To determine the number of captured DMP $N_{cap}(t)$, in SS with Jupiter,
as a function of time
we model numerically a constant flow of scattered particles
with energy distribution $d N_s \propto v f(v) dv$ per time unit.
The injection, capture, evolution and escape of DMP is described by 
the dark map (\ref{eq1}) with corresponding values of
scattered parameters $q, \theta, \varphi$ 
and corresponding to them $F(x)$ function with 
the scattering DMP distribution $dN_s \propto  dq d w$
(we use $q \leq q_{max}=4r_p$ since above this value
$F_{max}$ is very small).

The scattering and evolution processes are followed during the whole 
life time of SS taken as $t_{S}=4.5 \cdot 10^9 years$.
The total number of DMP, injected during time $t_{S}$ in the whole
energy range $0 \leq |w| \leq \infty$, is $N_{tot} \approx 1.5 \cdot 10^{14}$
with $N_H =4 \cdot 10^9$ scattered DMP in the Halley comet range
$0 < |w| \leq  w_H \approx 0.005$
($\kappa=N_{tot}/N_H \approx 2u^2/(3v_p^2 w_H) \approx 3.8 \cdot 10^4$,
only DMP with $|w|<F_{max}$ participate in dynamics).
We used a random grid of initial $q, \theta, \varphi$ values
with up to $N_0=4 \cdot 10^5$ grid points
and $N_i$ injected orbits at each grid point
with $N_H=N_0 N_i$.

\begin{figure}
\begin{center}
\includegraphics[width=0.48\textwidth]{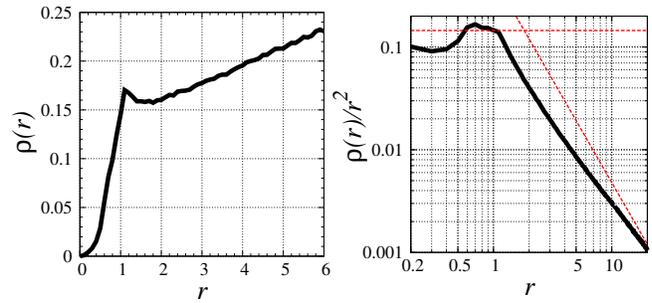}
\end{center}
\vglue -0.2cm
\caption{{\em Left panel}: Radial density $\rho(r) \propto dN/dr$
at  present time $t_S$ for SS with Jupiter
averaged over all $N_{AC}$ DMP 
(the normalization is fixed as $\int_0^{6r_p} \rho dr =1$,
$r_p=r_J=1$).
{\em Right panel}: Volume density 
$\rho_v=\rho/r^2$ from the data of left panel,
dashed line shows slope -2, horizontal line
shows average density for $r_p/5 \leq r \leq r_p$.
\label{fig4}}
\end{figure}

The time dependence $N_{cap}(t)$ in Fig.~\ref{fig3} shows that
initially it grows linearly with time. This growth slows down
after a time scale $t_d \approx 10^7$ years. For 
a finite  SS region with $w> 1/20$ we see that
there is a saturation of captured number of DMP. Indeed,
according to the results of \cite{halley} for the Halley
comet 
the scale $t_d \sim 10^7 yr$ is a typical scale of diffusive escape of 
a comet or DMP from SS due to chaotic diffusion in energy.
The analytical estimates given by \cite{halley,khriplovich2009}
also give a similar escape time. Thus after that 
time the injected flow is compensated by the escape process
and we obtain a system in an equilibrium state
with a fixed number of captured DMP with a certain 
energy distribution $\rho(w)$. 

An example of such a distribution
for a typical orbit parameters $q, \theta, \varphi$ at 
present time $t=t_S$ 
is shown in the inset of the right panel of Fig~\ref{fig3}
(gray curve). 
There is a peak of density
at small energies $w< 0.002$ where the orbital period
is very long and chaotization is slow. For 
the range $0.002 < w < w_{ch}$ 
we have an approximate algebraic decay
$\rho \sim 1/w^{3/2}$ which corresponds to the ergodic measure 
where DMP density is proportional to the orbital period
$T_w \sim 1/w^{3/2}$. The chaotic diffusion to large energies
is stopped by a critical invariant Kolmogorov-Arnold-Moser (KAM) 
curve which separates chaos region from integrable one
at $w=w_{ch}$.

The analytical estimate of \cite{khriplovich2009}, based on 
the Chirikov criterion (see \cite{chirikov1979,lichtenberg}), gives for 
$F_{max} \approx 5 m_p/M_S$ the value $w_{ch} \approx 0.3$
that is in a good agreement with the case of Fig.~\ref{fig3}
where $w_{ch} \approx 0.2$. In a region
$w_{ch}/2 < w < w_{ch}$ we have stability islands,
being well visible in the Poincar\'e section,
that gives significant fluctuations in density $\rho(w)$.
However, for $w<w_{ch}/2$ the chaos component is
homogeneous in the phase plane $(w,x)$. This means that
DMP are injected in the chaotic component of
a chaotic layer around separatrix $w=0$,
and thus the DMP dynamics in SS is essentially chaotic.

\section{Density and mass of captured dark matter}
\begin{figure}
\begin{center}
\includegraphics[width=0.48\textwidth]{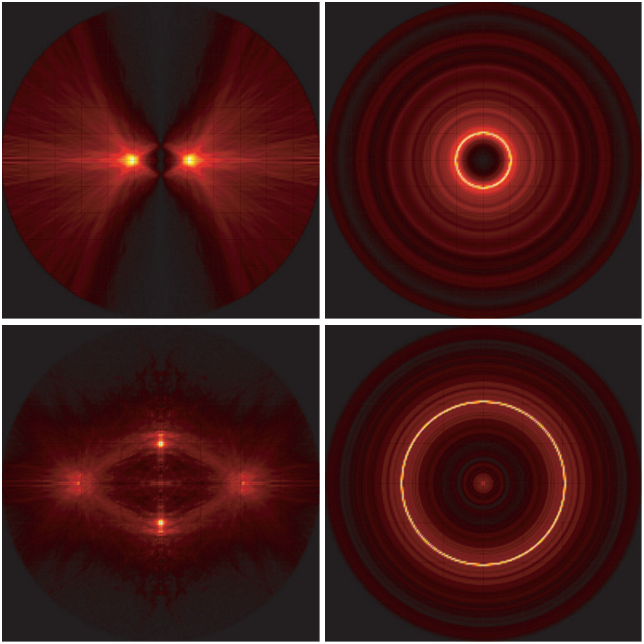}
\end{center}
\vglue -0.2cm
\caption{
Density of captured DMP in SS at present time $t_S$. 
{\em Top panels:} DMP surface density 
$\rho_s \propto dN/dzdr_\rho$ 
shown at {\em left}  
in cross plane $(0,y,z)$ perpendicular to Jupiter orbit
(data are averaged over $r_\rho=\sqrt{x^2+y^2}=const$),
at {\em right} in Jupiter plane $(x,y,0)$;
only the range $|r| \leq 6 r_J$ around the Sun is shown. 
{\em Bottom panels:} corresponding 
DMP volume density $\rho_v \propto dN/dxdydz$ at {\em left}
in plane  $(0,y,z)$, at {\em right} in Jupiter plane 
$(x,y,0)$;
only the range $|r| \leq 2 r_J$ around the Sun is shown. 
Color is proportional to density
with yellow/black for maximum/zero density.
\label{fig5}}
\end{figure}

To obtain DMP space density we consider $N_{tot}$ 
scattered orbits as described above.
Their time evolution is described by
the dark map (\ref{eq1}) 
up to the present moment of time $t_S$. We
keep in memory the initial
orbit parameters $q, \theta, \varphi$
of captured orbits.
Then we consider only those with
$w>4 \cdot 10^{-5}$  
during the time interval $\delta t_S/t_S= \pm 10^{-3}$ 
near time moment $t_S$ collecting
$\delta N_{AC} \approx 6.2 \cdot 10^6$ orbits
(while instantaneously we have $N_{AC} \approx 3.3 \cdot 10^5$).
For these $\delta N_{AC}$ DMP their dynamics in real 
space is recomputed from their values of
  $q, \theta, \varphi, w, x$ during the time period of
$\Delta t \approx 100$ Jupiter orbital periods using Newton equations.

The radial density $\rho(r)$ of DMP is obtained
by averaging over $10^3$ points randomly and homogeneously
distributed over this time interval $\Delta t$ for each of
$\delta N_{AC}$ orbits.
The obtained normalized
radial distribution $\rho(r)$ in shown in Fig.~\ref{fig4}
with the corresponding average volume density $\rho_v=\rho/r^2$. 
It corresponds to a stationary equilibrium regime appearing
at $t \gg t_d$ when injection and escape flows compensate each other.
The striking feature of the obtained result
is that for $r>r_p$ we find $\rho(r) \approx const$, This means that
the total DMP mass in a radius $r$ grows linearly with $r$.

According to the virial theorem  such a profile
gives a velocity of visible matter independent of radius
$v_m^2 \sim \int_0^r \rho(r') dr'/r \sim \rho(r)$, 
being similar to those found in galaxies 
when the DMP mass is dominant  as discussed by \cite{rubin,bertone2005}.
Another important feature is that the DMP volume density $\rho_v$
remains approximately constant for $r<r_p=r_J$. However,
for $r > r_J$ this density drops as inverse square distance from the Sun.
Thus we find that a simple model of SS with one rotating planet
is able to reproduce significant features of observed
DMP density distribution in galaxies.

Let us note that the radial density 
$\rho(r) \propto dN/dr$ is only approximately 
constant for $r>r_p$. Indeed, a 
formal fit of data of Fig.~\ref{fig4} right panel
in the range $2 < r/r_p<20$ gives
$\rho_v \sim 1/r^{\beta}$ with $\beta=1.53 \pm  0.002$.
We can argue that this dependence can be understood 
from  the ergodic measure of effectively 
one-dimensional chaotic radial dynamics: 
$d \mu \sim dN \sim \rho dr \sim \int dt dw (dN/dw)  \sim dt \sim dr/v_r 
\sim \sqrt{r} dr$ 
(assuming that $dN/dw$ is peaked near $w \approx 0$
as it is seen in the inset of Fig.\ref{fig3} and 
hence the radial velocity
$v_r \sim \sqrt{1/r-w} \sim 1/\sqrt{r}$ and $\rho_v \sim 1/r^{3/2}$).
Such a dependence would lead to velocity of visible matter
$v_m \propto \sqrt{\rho} \propto r^{1/4}$
if the DMP mass would be dominant, as it is the case in
galaxies as discussed by \cite{rubin}.

In fact the data presented by    \cite{rubin}
(see Fig.7 and Eqs.(1,2) there) 
are compatible with the dependence $v_m \propto r^{0.35}$
which is close to the above theoretical estimate.
However, in SS the DMP mass is small compared to the visible matter
and hence the case of galaxies should be analyzed in 
more detailed way using self-consistent conditions for DMP
distribution which would modify the second equation 
in the dark map. 
Though the above arguments can be useful for 
analysis of DMP distribution at
$r \gg r_p$, in this work 
we perform the density analysis mainly
inside the Neptune orbit where the radial density $\rho(r)$
can be considered as  approximately constant.

From the data of Figs.~\ref{fig3}
we determine the fraction $\eta_{AC} =N_{AC}/N_{tot} \approx 2.2 \cdot 10^{-9}$ 
of DMP captured at time $t_S$ at energies $w>4 \cdot 10^{-5}$
and related fraction $\eta_{20} \approx 1.5 \cdot 10^{-11}$
at energies $w>1/20$. From Fig.~\ref{fig4} we determine the
fraction of $N_{AC}$ orbits in the volume $r \leq 6 r_p$
with $\eta_{r6} \approx 4.3 \cdot 10^{-4}$ and in the volume
$r \leq r_p$ with $\eta_{r1} \approx 2.3 \cdot 10^{-5}$. 
The DMP mass corresponding to these fractions is obtained by
multiplication of these fractions by the total mass 
of DMP flow passed in the corresponding range $q \leq 4 r_p$: 
$M_{tot}= \int_0^{\infty} dv v f(v) \sigma \rho_g t_S 
\approx  69 \rho_g t_S k r_p M_S/u \approx 0.9 \cdot 10^{-6} M_S$
where we use the cross-section $\sigma = \pi r_d^2 = 8\pi k M_S r_p/v^2$
for injected orbits with $q \leq 4 r_p$, $k$ is the gravitational constant
($u/v_p \approx 17$). 
Thus the mass of DMP with $w > 4 \cdot 10^{-5}$
is $M_{AC} \approx \eta_{AC} M_{tot} \approx 2 \cdot 10^{-15}M_S$,
and in a similar way the mass at $w>1/20$ is 
$M_{20} \approx \eta_{20} M_{tot} \approx 1.3 \cdot 10^{-17} M_S$.
The mass $M_{AC}$ can be estimated as a mass of DPM with 
$|w|<w_H$ absorbed by $F \sim \sin x$ kick 
during the diffusion time $t_d$ that
gives $M_{AC} \sim v_p^2 w_H t_d M_{tot}/(\pi u^2 t_S) \sim 10^{-8} M_{tot}
\sim  10^{-14} M_S$ being only by 
a factor 5 larger the above numerical value. 

The mass of DMP in the volume of Neptune orbit radius $r < 6r_p$ is
$M_{r6} = \eta_{r6} M_{AC} \approx 0.9 \cdot 10^{-18} M_S 
\approx 1.7 \cdot 10^{15}g$
and in the radius $r < r_p$ the DMP mass is
$M_{r1} = \eta_{r1} M_{AC} \approx 4.6 \cdot 10^{-20} M_S 
\approx  10^{14} g$. The average volume density 
of captured DMP inside the Jupiter orbit  sphere $r<r_p=r_J$
is $\rho_{J} = 3 M_{r1}/(4\pi r_p^3) \approx 1.2 \cdot 10^{-4} \rho_g 
\approx 5 \cdot 10^{-29} g/cm^3$. 
Thus, the density of captured DMP is much smaller
than the galactic DMP density. However, it is by a factor $4 \cdot 10^3$
larger than the equilibrium DMP galactic density
$\rho_{gH} \approx 0.25 \rho_g/\kappa^{3/2} \approx 1.4 \cdot 10^{-32}g/cm^3$
taken in the energy range $0<|w|<w_H$.

The density distribution of captured DMP in SS is shown in Fig.~\ref{fig5}.
We see that the density decreases with $r$ at $r> r_J$ in agreement with 
Fig.~\ref{fig4}. A characteristic bulge is formed around the Jupiter orbit.
A maximal local volume density is about 10 times larger than the average
density $\rho_J$ inside $r<r_J$. 

\section{Discussion}
For further studies it is desirable to take into account the contribution
of other planets even if the results presented by \cite{halley}
show that the main features of the dynamics are well described 
only by Jupiter contribution considered here.
It is natural to expect that, as in the SS with one planet, 
the DMP dynamics in galaxies is dominated by a few stars 
rotating around the central black hole and thus a constant
radial DMP density behind Jupiter orbit found here 
should be typical for
such galaxies in agreement with observational data 
discussed by \cite{rubin}.

In global our studies show that the average captured DMP density inside
the Jupiter orbit is by a factor $10^4$ smaller than the galactic DMP 
density $\rho_g$. The main reason for that is a small value of captured
DMP energy $w_{cap} \sim m_p/M_S \sim 0.005$
 which is very small compared to the dimensional galactic
DMP velocity $u/v_p \sim 17$.  However, if we consider
the galactic density in the capture energy range of $0<|w|<w_{cap}$
than we find that it is significantly enhanced by a factor $4\cdot10^3$ 
due to the capture process considered here.
Thus, the further analysis of chaotic capture process
of dark matter in binary systems can bring interesting results.

It would be also interesting to consider the inverse ionization process of
DMP. According to the dark map (\ref{eq1}) the escape velocity square
of DMP from a binary system of a star of mass $m_s$ rotating
in a vicinity of a black hole of
mass $M_b$ is  $v_d^2 \sim (m_s/M_b) v_s^2$. For a star moving in a vicinity
of  the Schwarzschild radius we may have the star velocity $v_s \sim c/3$
and for the mass ratio $m_s/M_b \sim 0.01$ we obtain
the escape velocity of DMP $v_d \sim c/30 \approx 10^4 km/s$ that is almost 
hundred times larger than the average galactic DMP velocity $u \sim 200km/s$.
Any other body of mass significantly smaller $m_s$ is ejected with a similar
velocity that can generate  compact wandering black holes 
crossing  the universe at high velocity $v_d$.
Thus, the stars on a distance of Schwarzschild radius from black holes
can work as some kind of black hole accelerators generating
high velocity DMP in the universe.

We thank I.B.Khriplovich for useful discussions. 
A part of numerical computations has been performed 
at the m\'esocentre de calcul de Franche-Comt\'e.


\label{lastpage}

\end{document}